\documentclass[submission,copyright,creativecommons]{eptcs}
\providecommand{\event}{QAPL 2016} 
\usepackage{fancyhdr}
\usepackage{amsmath,amsthm,amssymb}
\usepackage{graphicx}
\usepackage{hyperref}
\usepackage{listings}
\usepackage{subfigure}
\usepackage{multirow}
\usepackage{tikz}
\usetikzlibrary{automata,positioning}
\usepackage[absolute,overlay]{textpos}
\usetikzlibrary{shapes,fit}
\usepackage{algorithm}
\usepackage{algorithmic}
\usepackage{url}

\newcommand{\loc}{\ensuremath{\mathbf{\ell}}}

\newcommand{\redtext}[1]{\textcolor{black}{#1}}

\title{Location Aggregation of Spatial Population CTMC Models}
\author{Luca Bortolussi
\institute{University of Trieste, Trieste, Italy}
\institute{CNR-ISTI, Pisa, Italy}
\email{luca@dmi.units.it}
\and
Cheng Feng
\institute{University of Edinburgh\\
Edinburgh, UK}
\email{s1109873@sms.ed.ac.uk}
}
\def\titlerunning{Aggregation of Spatial PCTMC Models}
\def\authorrunning{L. Bortolussi \& C. Feng}
\begin{document}
\maketitle

\begin{abstract}
In this paper we focus on spatial Markov population models, describing the stochastic evolution of populations of agents, explicitly modelling their spatial distribution, representing space as a discrete, finite graph. More specifically, we present \redtext{a} heuristic approach to aggregating spatial locations, which is designed to preserve the dynamical behaviour of the model whilst reducing the computational cost of analysis. Our approach combines stochastic approximation ideas (moment closure, linear noise), with computational statistics (spectral clustering) to obtain an efficient aggregation, which is experimentally shown to be reasonably accurate on two case studies: an instance of epidemic spreading and a London bike sharing scenario. 
\end{abstract}

\section{Introduction}
Population processes are widely used to describe a large variety of systems, including systems in biological \cite{gillespie1977exact}, ecology \cite{allen2010introduction},  performance engineering \cite{stefanek2011fluid},  smart cities facilities like bike sharing \cite{fricker2014incentives,guenther2013journey}, and the spreading of epidemics \cite{andersson2000}. Most of these systems are spatially distributed, and space has to be explicitly modelled to properly capture the relevant features of their dynamics. For example, interactions may only be allowed for entities which are co-located or within a certain physical distance of each other,
or space may be segmented in such a way that even physically close entities are unable to communicate or interact.  Furthermore, movement in space can be a crucial aspect of the behaviour of entities within the system.  Epidemic spreading, in which infection can pass only by physical contact, is a clear example in this sense \cite{andersson2000}.  Bike sharing is another one, as the geographical location of bike stations influences the travelling time, and different stations have different demands for acquisition or deposit of bikes at different times of the day \cite{fricker2014incentives,guenther2013journey}. 

Population processes are often modelled as Markovian stochastic processes, mostly as a  subclass of continuous-time Markov chains (CTMCs)  known as Population CTMCs (PCTMCs).  These models have a very large or even infinite state space, a fact resulting in a lot of work  in the computer science community  to craft efficient algorithms for their analysis, ranging from specialised stochastic simulation to the use of mean-field \cite{tutorial} and moment closure techniques \cite{hasenauer2013,guenther2013moment}. These methods, in particular, approximate the large or infinite set of linear Kolmogorov equations by a much smaller set of non-linear differential equations, capturing the mean, variance, and possibly other higher order moments. 

Spatial extensions of PCTMCs typically introduce discrete representations of space in terms of locality,  connecting them in a general topology represented by a (weighted) graph \cite{valuetools2014}. One of the effects of explicit modelling of space is the increase of the computational cost of analysis of the system. For instance, a model with $l$ localities will increase the number of equations for the variance in any moment closure approach by a factor of $O(l^2)$. 

However, although space shapes the behaviour of the system, our interest is often in space-free properties, like the total number of infected individuals in an epidemic scenario, or in localised properties, like the number of available bikes in a given station or geographic area. In these cases, a full representation of space may not be needed to compute such quantities with a reasonable accuracy, and model simplification and abstraction can be a viable strategy. 

In this paper, we pursue this direction by developing an efficient approach to aggregate locations showing a similar \redtext{dynamical} profile. This operation will preserve the dynamical behaviour of the model, but lower the computational cost: for instance, if the number of locations decreases from $l$ to $k$, then the number of differential equations for the variance is reduced by a factor $O(k^2/l^2)$. In particular, here we will focus on the reduced cost of the analysis of the full stochastic model by using standard simulation algorithms  \cite{gillespie1977exact}, as this is the most expensive but also most informative computational technique (excluding numerical integration of the Kolmogorov equations, which for PCTMC is unfeasible due to state space explosion). 

Our approach is based on aggregating locations using state-of-the-art spectral clustering approaches, based on metrics between locations that take the steady state mean or distribution into account. Rather than working with exact solution or estimations of these quantities by simulation, which would be computationally expensive, we obtain them applying stochastic approximation ideas, either by solving mean-field equations for the mean or using a linear-noise-like approximation of the distribution, obtaining mean and variance from moment closure equations. 
The feasibility, effectiveness, and accuracy of our method is discussed on two case studies: an epidemic model and a London bike sharing scenario.

The paper is organised as follows: Section \ref{sec:PCTMC} introduces the formalism of spatial PCTMCs we will use afterwards and briefly introduces exact and approximate analysis techniques. Section \ref{sec:aggregation} discusses in detail our aggregation approach, and Section \ref{sec:cases} presents the two case studies. Conclusions are drawn in Section \ref{sec:conclusion}.

\section{Spatial PCTMC Models}
\label{sec:PCTMC}
A Population Continuous Time Markov Chain (PCTMC)  is a Markovian stochastic process evolving in continuous time. It consists of a number of individually indistinguishable interacting agents of different types which can be in different internal states, so that the state of the system can be described by counting how many agents of each kind are in the system.  
Agents' interactions are described by a set of transitions, which will change the internal state of one or more agents. 
In this paper, we specifically consider \emph{spatial} PCTMC models in which agents are distributed in a finite set of discrete locations, with interactions typically happening either in the same location or in neighbouring ones. Formally, a spatial PCTMC can be expressed as a tuple $\mathcal{P}=(\mathcal{L},\mathbf{X}, \mathcal{T})$:
\begin{itemize}
\item $\mathcal{L}=(\loc_1,...,\loc_l)$ is the set of discrete locations in the model, where $|\mathcal{L}|=l$ denotes the total number of locations.
\item $\mathbf{X}=(\mathbf{X}@\loc_1,...,\mathbf{X}@\loc_l)\in \mathbb{Z}_{\geq 0}^{N}$ is an integer vector representing the agent populations distributed over all the locations in the model, where $|\mathbf{X}|=N=l\times n$ is the total number of distinct agent populations in the model with $n$ \redtext{representing} the number of agent types. $\mathbf{X}@\loc_i$ is an sub-vector with the $j$th component, $X_j@\loc_i$ representing the current population of the agent type $j$ at location $i$.  We use $\mathbf{X_0}$ to denote the initial state of the model. 
\item $\mathcal{T}=\{ \tau_1,...,\tau_m \}$ is the set of transitions with size $|\mathcal{T}|=m$, of the form $\tau=(r_{\tau}(\mathbf{X}),\mathbf{D}_{\tau})$, where:
        \begin{enumerate}
            \item $r_{\tau}(\mathbf{X}) \in \mathbb{R} \geq 0$ is the rate function, associating with each transition the rate of
an exponential distribution, depending on the global state of the model.
            \item $\mathbf{D}_{\tau} \in \mathbb{Z}^N$ is the update vector which gives the net change for each element of $\mathbf{X}$ caused by transition $\tau$. Similarly, we let $\mathbf{D}_{\tau}@\loc_i$ denote the update vector of agent populations at location $i$.
		\end{enumerate}
\end{itemize}
Transition rules can be easily visualised in the chemical reaction style, as
\begin{eqnarray*}
u_{\tau}^1@\loc_1  + \ldots + u_{\tau}^n@\loc_l  &\longrightarrow &   v_{\tau}^1@\loc_1  + \ldots + v_{\tau}^n@\loc_l  \quad \text{at rate $r_{\tau}(\mathbf{X})$}
\end{eqnarray*}
where  the net change of agents of type $i$ in location $\loc_j$ due to transition $\tau$ is given by  $ D_{\tau}^i@\loc_j = v_{\tau}^i@\loc_j - u_{\tau}^i@\loc_j$  ($1\leq i \leq n$, $1\leq j\leq l$). This is a general format, encompassing local interactions, interactions between agents in neighbouring locations, and movement in space. 
The tuple $\mathcal{P}$ contains all the information that is needed to build a CTMC on the state space $\mathbb{Z}_{\geq 0}^{N}$: its infinitesimal generator $Q$ is given by $Q(\mathbf{x},\mathbf{y}) = \sum\{r_{\tau}(\mathbf{x}) \vert \mathbf{D}_\tau =  \mathbf{y}- \mathbf{x}\}$, for $\mathbf{x}\neq \mathbf{y}$. Spatial PCTMC can be simulated using standard stochastic simulation algorithms \cite{gillespie1977exact}.

\subsection{Moment Closure, Mean Field, and Linear Noise}
\label{sec:mf}
The analysis of the stochastic model underlying a spatial PCTMC is not an easy task. Numerical methods are hindered by the state space explosion, and even stochastic simulation suffers from the presence of localities and large populations. Furthermore, it is known that for large populations the behaviour of the stochastic model becomes deterministic and converges to the solution of a the mean field differential equation \cite{tutorial,kurtz1981}, which in the formalism defined above takes the form
\[
\mathbf{\dot{X}} = F(\mathbf{X}), \qquad \text{with }  F(\mathbf{X})= \sum_{\tau} \mathbf{D}_\tau r_{\tau}(\mathbf{X})
\]
If populations in each location are in the order of tens or hundreds of thousands, mean-field equations are generally very accurate \cite{tutorial}. However, when populations are smaller, in the order of hundreds, stochastic fluctuations still have an important role, and the linear-noise  (central limit) approximation performs better \cite{kurtz1981,vankampen}. The idea behind linear noise is to approximate the original stochastic model by a linearized Markov process in continuous \redtext{states}, whose solution is a Gaussian process. The distribution of this process at a given time is thus characterized by  solving equations for the mean (incidentally, the mean-field equations above) and equations for the covariance. 

An alternative strategy for the analysis of the spatial PCTMC is to derive equations for the moments of the population variables \cite{hasenauer2013,guenther2013moment}, up to order $k$. Due to non-linearities in the rates, there is no exact closed form of these equations, and the differential equations for moments of order $k$ depend on higher-order ones. Hence, equations are closed by relying to some heuristic \cite{hasenauer2013,guenther2013moment}. Typically, moment closure equations give a better estimation of mean and variance than linear noise, as knowledge of higher-order moments introduce correction terms in the equations of lower order ones. The first two moments can still be used to build a Gaussian approximation to the true distribution at a given time $t$, formally invoking a maximum entropy argument \cite{andreychenko2015,lanciani2014}.

\section{Aggregation of Locations of Spatial PCTMC}
\label{sec:aggregation}
In this section we present the computational methodology to reduce a spatial model by aggregating locations. The main motivation underlying this approach, as discussed in the introduction, is to reduce the computational effort in the analysis of the model. This effort, in fact, is proportional to the number of locations. This is the case for stochastic models, which need to simulate and keep track of the state of populations in each location, and for approximate analysis techniques like moment closure. In fact, even when considering only equations for the  second order moments, the number of such equations grows quadratically with the number of locations. However, if our final goal is to capture the behaviour of the system at the global level (i.e. to know how many agents of each kind are in the system), then dealing with the full set of locations may incur excessive work which can be reduced by grouping together locations having a similar overall behaviour. To achieve this goal while keeping the error low, we propose the following heuristic scheme, sketched here and detailed in the following subsections.
\begin{enumerate}
\item Define distance metrics between locations taking into account the dynamical behaviour of the system (Section \ref{sec:distance}). More specifically, we want  locations showing a similar steady state behaviour to be  clustered together. 
We will define two distances, described below in increasing order of precision:   
\begin{enumerate}
\item Mean field distance. We will consider the distance between the mean of populations at steady state, approximated by first order moment closure, which corresponds to the mean field abstraction of the PCTMC model (see Section \ref{sec:mf}).
\item Linear noise distance. We will consider the distance between a Gaussian approximation of the steady state distribution, computed from moment closure equations for mean and variance.  
\end{enumerate}
\item Cluster the locations using the previously computed distance  (Section \ref{sec:clustering}). We will use a spectral clustering algorithm on graphs, exploiting the eigengap heuristic to identify the number of clusters. 
\item Given a clustering of locations,  construct the reduced model by suitably aggregating together the PCTMC transitions and variables  (Section \ref{sec:reduction}).
\end{enumerate}

%
%

\subsection{Distance between Locations}
\label{sec:distance}
We will consider distances between populations of different locations at a given time. The choice of this time is important, and should not be taken too small, in order to minimise the effect of the  initial state of the model. In fact, if we  took information about the whole trajectory into account, different initial conditions between the populations of two locations would contribute to the distance, often resulting in a separation between the two locations.  However, especially if we consider aggregated quantities, like the total number of agents of a certain kind across all locations, this difference is not very relevant, and better results can be obtained by comparing the behaviour after a finite but large time. We will see the experimental validation of this choice in the next Section. Note that we do not consider steady state behaviour (though for a very  large time steady state would be reached), as we will  use mean-field or linear noise approximations, which do not necessarily converge at steady state. 
%
%
We will consider two distances between each pair of locations $\loc_i$ and $\loc_j$, with increasing levels of accuracy:
\begin{enumerate}
\item \emph{Mean field distance} $d_E(\loc_i,\loc_j)$. This is just the Eucliden distance between the average value of populations of locations $\loc_i$ and $\loc_j$ at steady state, i.e.  $d_E(\loc_i,\loc_j) = \| \mu_{\mathbf{X}@\loc_i}-\mu_{\mathbf{X}@\loc_j }\|$, where $\mu_{\mathbf{X}@\loc}$ is the mean of the population (vector) $\mathbf{X}@\loc$ at steady state.
To compute this distance efficiently, we resort to the mean-field approximation or to any first order moment closure, see Section \ref{sec:mf}. This reduces the problem to the numerical integration of $N$ differential equations ($N=n\times l$), until they reach equilibrium (or until their temporal average stabilise in case they oscillate).\footnote{We are implicitly assuming that the mean field or moment closure equations will not show chaotic behaviour, which is typically the case for models of interacting agents. }
\item \emph{Linear noise distance} $d_L(\loc_i,\loc_j)$. To capture more accurately the steady state behaviour, we can consider a distance between the full distributions. To this end, we will resort to the Bhattacharyya distance $d_B(Y_1,Y_2)$ \cite{bhattachayya1943measure} which measures the distance of the distribution of two one-dimensional  variables $Y_1,Y_2$. more specifically, we will compute the Bhattacharyya distance between each popoulation k in locations $\loc_i$ and $\loc_j$, denoted by $d_B(X_k@\loc_i, X_k@\loc_j)$, and aggregate over all populations by taking the average: 
\begin{equation}
d_L(\loc_i, \loc_j) = \frac{1}{n} \left(d_B(X_1@\loc_i, X_1@\loc_j) + \cdots + d_B(X_n@\loc_i, X_n@\loc_j)\right)  \nonumber 
\end{equation}
In order to compute $d_B(X_k@\loc_i, X_k@\loc_j)$ without the need to estimate the full distribution, we make a linear noise assumption, i.e. that the steady state distribution is approximately Gaussian, see Section \ref{sec:mf}.   Under this hypothesis,  $d_B(X_k@\loc_i, X_k@\loc_j)$ can be calculated by the following equation:
             \begin{equation}
               D_B(X_k@\loc_i, X_k@\loc_j)={\frac {1}{4}}\ln \left({\frac {1}{4}}\left({\frac {\sigma_{X_k@\loc_i}^{2}}{\sigma _{X_k@\loc_j}^{2}}}+{\frac {\sigma _{X_k@\loc_j}^{2}}{\sigma _{X_k@\loc_i}^{2}}}+2\right)\right)+{\frac {1}{4}}\left({\frac {(\mu _{X_k@\loc_i}-\mu _{X_k@\loc_j})^{2}}{\sigma _{X_k@\loc_i}^{2}+\sigma _{X_k@\loc_j}^{2}}}\right) \nonumber
             \end{equation}
             where $\mu_{X@\loc}$ and $\sigma_{X@\loc}$ denote the mean and variance of the population variable $X@\loc$ at the steady state. To numerically compute the values of $\mu_{X@\loc}$ and $\sigma_{X@\loc}$, we resort to the  normal moment-closure  approximation of \cite{guenther2013moment}, which can be obtained at a much lower computational cost than by simulating the PCTMC, by integrating $O(N^2)$ differential equations.
\end{enumerate}
We observe  that the  cost of computing $d_L(\loc_i, \loc_j)$ is significantly higher than the cost of computing $d_E(\loc_i, \loc_j)$, as we need to integrate $O(N^2)$ differential equations rather than $O(N)$. However, this cost is balanced by a higher accuracy in the reduced system, though for certain models (essentially those having similar variance for the same agent kind at different locations in a cluster) accuracy is comparable. In practice, $d_E$ should be used when the cost of solving moment closure equations for the variance is very high due to the very large number of locations. Hence, aggregation with respect to $d_E$ can be seen as a reduction of the number of moment closure equations. When the goal is to reduce the cost of stochastic simulation, and solving moment closure equations for the variance is cheap, it is better to rely on the metric $d_L$. \redtext{In general, our method should be applied when stochastic simulation of the spatial PCTMC model requires excessive computational time, and computing $d_E$ or $d_L$ is much cheaper than the cost of stochastic simulation.}

\subsection{Spectral Clustering}
\label{sec:clustering}
Spectral clustering methods are common graph-based approaches to (unsupervised) clustering of data \cite{von2007tutorial}. The dataset is composed of $n$ objects  $S = {s_1,\ldots, s_n }$, among which some local symmetric and non-negative similarity measure $A_{i,j}$ is defined. $A_{ij}$ is often obtained from a distance or difference measure $d(s_i,s_j)$ between the objects. This information is then arranged in a weighted graph $G = (S,A)$.  Within this framework, clustering is translated into a graph partitioning problem. The most common class of spectral approaches for graph partitioning (in $k$ subsets) is to map the original data into the first $k$ eigenvectors of some normalized version of the similarity matrix $A$ and then apply a standard clustering algorithm such as k-means on these new coordinates.

Among the most commonly used spectral clustering algorithms are the unnormalized spectral clustering \cite{von2007tutorial}, normalized spectral clustering according to Shi and Malik \cite{shi2000normalized}, and normalized spectral clustering according to Ng, Jordan, and Weiss \cite{ng2002spectral}. The three algorithms are very similar; their main distinguishing feature is the fact that they use three different graph Laplacians. As an illustration, the normalized spectral clustering algorithm according to Ng, Jordan, and Weiss is given as follows.
Given a set of objects $S = {s_1,\ldots, s_n }$ that we want to cluster into $k$ subsets:
\begin{enumerate}
\item  Form the similarity matrix $A \in \mathbb{R}^{n\times n}$ defined by $A_{ij} = \exp(-\frac{d(s_i-s_j)^2}{2\sigma^2})$ if
$i \neq j$, and $A_{ii} = 0$.
\item  Define $D$ to be the diagonal matrix whose $(i, i)$-element is the sum of $A$'s $i$-th row, and construct the normalized Laplacian matrix $L = I - D^{-1/2} A D^{-1/2}$
\item  Find $U_1 , U_2 , ... , U_k$ , the first $k$ eigenvectors of $L$ with $k$ smallest eigenvalues,  and form the matrix $U =[U_1,U_2,...,U_k] \in \mathbb{R}^{n\times k}$ by stacking the eigenvectors in columns.
\item  Form the matrix $Z$ from $U$ by renormalizing each of $U$'s rows to have unit length
(i.e. $Z_{ij} = U_{ij}/(\sum_j U_{ij}^2)^{1/2}$).
\item  Treating each row of $Z$ as a point in $\mathbb{R}^k$ , cluster them into $k$ clusters via k-means.
\item  Finally, assign the original object $s_i$ to cluster $j$ if and only if row $i$ of the matrix $Z$ was assigned to cluster $j$. 
\end{enumerate}
We will use the above algorithm hereafter for the clustering of locations in spatial PCMTCs .
\subsubsection{Application to Aggregation of Locations of spatial PCTMC}
Given a spatial PCTMC model we wish to aggregate its locations $\mathcal{L}=(\loc_1,...,\loc_l)$. The first step is to compute one of the two distances $d_E$ or $d_L$ of the previous section. From this metric, we can derive the similarity matrix $A \in \mathbb{R}^{l\times l}$ for the locations $\mathcal{L}$ by using Gaussian kernel with width $\sigma$:
\begin{equation}
A_{ij} = \exp \left(-\frac{d_\star(\loc_i, \loc_j)^2}{2\sigma^2}\right) \nonumber 
\end{equation}
Then, the standard spectral clustering algorithms can be applied to cluster the locations with a specific choice of the number of clusters  $k$. We will refer to the locations' clusters by $\hat{\mathcal{L}}=(\hat{\loc}_1,...,\hat{\loc}_k)$.

In order to select $k$, we rely on the eigengap heuristic \cite{moharbojan}. Specifically, we choose the number of clusters $k$ such that all eigenvalues $\lambda_1,\ldots,\lambda_k$ are very small, but $\lambda_{k+1}$ is relatively large. A realisation of this heuristic will be shown while discussing case studies.

\subsection{Model Reduction}
\label{sec:reduction}
In this section, we show how to generate a reduced version of a spatial PCTMC model once the locations $\mathcal{L}$ in the original model have been clustered into the aggregated locations $\hat{\mathcal{L}}$. Suppose we have clustered locations in a spatial PCTMC into $k$ subsets. Formally, we want to map the original spatial PCTMC model $\mathcal{P}=(\mathcal{L},\mathbf{X}, \mathcal{T})$ to a reduced one  $\hat{\mathcal{P}}=(\hat{\mathcal{L}},\hat{\mathbf{X}},\hat{ \mathcal{T}})$, where $|\hat{\mathcal{L}}|=k < l=|\mathcal{L}|$, $|\hat{\mathbf{X}}|<|\mathbf{X}|$, and $|\hat{\mathcal{T}}| \leq  |\mathcal{T}|$. Hence, we need  to construct both the aggregated vector of  the populations of agents and the reduced set of transitions.

\subsubsection{Generating the aggregated vector of agent populations}
The algorithm for generating the aggregated vector of agent populations is fairly straightforward. The basic idea is to treat agents of the same type in the same cluster of locations as identical agents. Thus, we only need to sum up the populations of those identical agents. Algorithm~\ref{alg-1} gives the corresponding pseudo code for generating the aggregated vector of agent populations. Note that in the pseudo code, we use the same notations for the original spatial PCTMC model as in Section 2.

\begin{algorithm}
\begin{small}
  \caption{The Algorithm for generating the aggregated vector of agent populations}
  \label{alg-1}
  \begin{algorithmic}[1]
  	\REQUIRE $\mathbf{X}$, $\hat{\mathbf{X}}$, $\mathcal{L}$, $\hat{\mathcal{L}}$
  	\ENSURE $|\hat{\mathbf{X}}| = n \times k \wedge \hat{X}_i@\hat{\loc}_j=0 \quad \forall \hat{X}_i@\hat{\loc}_j \in \hat{\mathbf{X}} \quad $ \COMMENT{all elements in $\hat{\mathbf{X}}$ are set to zero initially}
    \FOR{$i=1$ \TO $l$}
     	\FOR{ $j=1$ \TO $c$ }
     		\IF{$\loc_i$ belongs to cluster $j$} 
     			\STATE{let $\hat{\mathbf{X}}@\hat{\loc}_j = \hat{\mathbf{X}}@\hat{\loc}_j + \mathbf{X}@\loc_i$}
     		\ENDIF
     	\ENDFOR 
     \ENDFOR
   \RETURN  $\hat{\mathbf{X}}$	
  \end{algorithmic}
  \end{small}
\end{algorithm}

\subsubsection{Generating the reduced set of transitions}
Three steps are taken to generate the reduced set of transitions. The first step is to copy the transitions in the original spatial PCTMC model to $\hat{\mathcal{T}}$. Meanwhile, the update of agent populations in the transition should be replaced by update of corresponding aggregated populations, the agent populations appearing in the rate function should also be replaced accordingly. The first step may generate many redundant transitions in which there is no agent population being updated. Thus, the second step is to remove redundant transitions from $\hat{\mathcal{T}}$. The last step is de-duplication, in which we combine all transitions with the same update vector into one transition. Algorithm~\ref{alg-2} gives the pseudo code for the three steps,  in which we use $|\hat{\loc}_i|$ to denote the number of locations belonging to cluster $\hat{\loc}_i$.  Note that this construction will produce an approximate model with respect to the original one, the reason being the treatment of rates. Approximation stems from the fact that in the rates of the aggregated  model we replace all occurrences of  each $X@\loc$ with $X@\hat{\loc}/|\hat{\loc}|$, where $\loc\in\hat{\loc}$. Hence we assume that for each $\loc_i,\loc_j\in\hat{\loc}$,  $X@\loc_i = X@\loc_j$. However, the heuristics we use to construct the aggregated model  guarantee that this condition should be roughly satisfied (at steady state).

%

\begin{algorithm}[t]
	\begin{small}
  \caption{The Algorithm for generating the reduced set of transitions}
  \label{alg-2}
  \begin{algorithmic}[1]
  	\REQUIRE $\mathcal{L}$, $\hat{\mathcal{L}}$, $\mathcal{T}$, $\hat{\mathcal{T}}$
  	\ENSURE $|\hat{\mathcal{T}}| = 0$
  	\FORALL{$\tau$ in $\mathcal{T}$}
  		\STATE{create a new $\hat{\tau}$, set $r_{\hat{\tau}} = r_{\tau}$, $|\mathbf{D}_{\hat{\tau}}| = n\times k \quad$ \COMMENT{all elements in $\mathbf{D}_{\hat{\tau}}$ are set to zero initially}}
     \FOR{ $i=1$ \TO $l$ }
     		 \FOR{ $j=1$ \TO $k$ }
     			\IF{$\loc_i$ belongs to cluster $j$} 
     			\STATE{let $\mathbf{D}_{\hat{\tau}}@\hat{\loc}_j = \mathbf{D}_{\hat{\tau}}@\hat{\loc}_j + \mathbf{D}_{\tau}@\loc_i$}
     		\ENDIF
     		\ENDFOR 
     	\ENDFOR 
     \FORALL{$X@\loc$ appears in $r_{\hat{\tau}}$}
     	 \FOR{ $i=1$ \TO $k$ }
     	 \IF{$\loc$ belongs to cluster $i$}
     	 \STATE{replace $X@\loc$ with $X@\hat{\loc}_i/|\hat{\loc}_i|$} 
     	 \ENDIF
     	 \ENDFOR
     \ENDFOR
     \ENDFOR
     \FORALL{$\hat{\tau}$ in $\hat{\mathcal{T}}$}
     \IF{All elements in $\mathbf{D}_{\hat{\tau}}$ equal zero}
     \STATE{remove $\hat{\tau}$ from $\hat{\mathcal{T}}$}
     \ENDIF
     \ENDFOR
     \FORALL{$\hat{\tau}_i$, $\hat{\tau}_j$ ($i \neq j$) in $\hat{\mathcal{T}}$}
     \IF{$\mathbf{D}_{\hat{\tau}_i} = \mathbf{D}_{\hat{\tau}_j}$}   
     \STATE{create a new $\hat{\tau}$, set $\mathbf{D}_{\hat{\tau}} = \mathbf{D}_{\hat{\tau}_i}$ and $r_{\hat{\tau}} = r_{\hat{\tau}_i} + r_{\hat{\tau}_j}$}
     \STATE{remove $\hat{\tau}_i$, $\hat{\tau}_j$ from $\hat{\mathcal{T}}$, add $\hat{\tau}$ to $\hat{\mathcal{T}}$}
	\ENDIF
	\ENDFOR   
   \RETURN  $\hat{\mathcal{T}}$	
  \end{algorithmic}
	\end{small}
\end{algorithm}

\section{Case Studies}
\label{sec:cases}
In this section we test our method on two case studies: a benchmark model of epidemic spreading, Section \ref{sec:SIS}, and a realistic model of a portion of the London bike-sharing system, Section \ref{sec:BS}.

\subsection{Spatial Epidemic Spreading Model}
\label{sec:SIS}
We first consider a classical epidemiological SIS model of individuals partitioned into $m$ communities, where individuals move between communities but infections only take place within communities. Each individual is considered to be susceptible (S) or infected (I) with respect to the disease. A continuous-time SIS epidemiological model is then applied to this population as follows: each individual, regardless of susceptible or infected, can move to his/her connected communities with a specific rate ($r_{ij}$). Each community is connected with three other randomly chosen communities.  Each infected individual can randomly make contact with a susceptible individual in the same community, and infect her with rate $\beta_i$ in community $i$. Finally, infected individuals independently recover to the susceptible state at rate $\mu$.

The model can be studied by a spatial PCTMC containing the following transitions:
\begin{align*}
S(\loc_i) &\rightarrow  I(\loc_i) & &\text{at $\beta_i \:\#(S(\loc_i)) \:\#(I(\loc_i))$} &\quad  & \forall i \in m\\
I(\loc_i) &\rightarrow   S(\loc_i) & &\text{at $\mu \:\#(I(\loc_i))$} &\quad  &\forall i \in m \\
S(\loc_i)  &\rightarrow   S(\loc_j) & &\text{at $r_{ij} \:\#(S(\loc_i))$} &\quad & \forall i,j \in m, \:  \text{$\loc_i$ and $\loc_j$ are connected}\\
I(\loc_i)  &\rightarrow   I(\loc_j) & &\text{at $r_{ij} \:\#(I(\loc_i))$} &\quad & \forall i,j \in m, \:  \text{$\loc_i$ and $\loc_j$ are connected}
\end{align*}where $S(\loc_i)$, $I(\loc_i)$ represent a susceptible, infected individual in Community $i$; $\beta_i$ represents the contact rate in community $i$; $r_{ij}$ denotes the rate for an individual to travel from Community $i$ to Community $j$.

\begin{figure}
  \centering
    \includegraphics[width=0.4\textwidth]{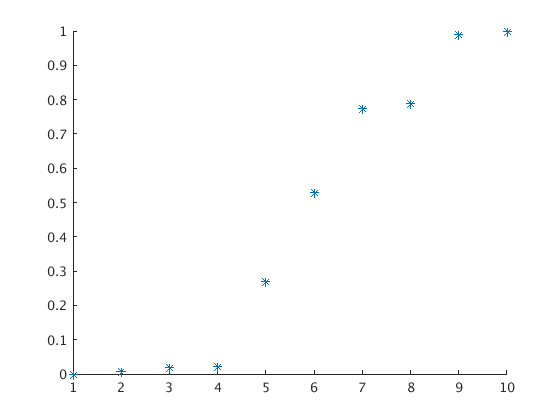}
        \includegraphics[width=0.4\textwidth]{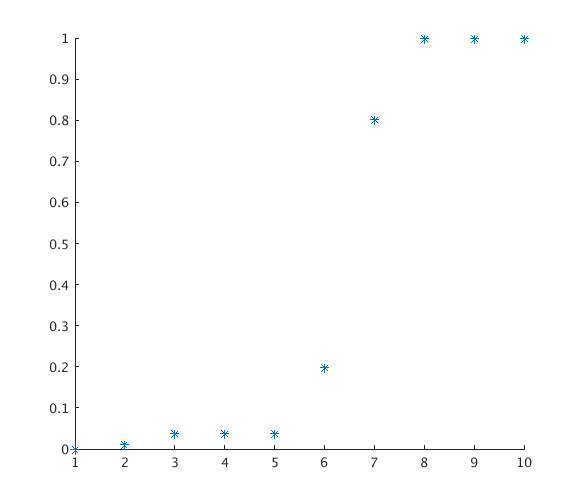}

    \caption{(left) The smallest 10 eigenvalues of the normalized Laplacian matrix of the SIS model. (right) The smallest 10 eigenvalues of the normalized Laplacian matrix of the bike-sharing model.}
    \label{fig:eigenvalues}
    \label{fig:eigenvaluesSIS}
\end{figure}

\begin{figure}
  \centering
    \includegraphics[width=0.45\textwidth]{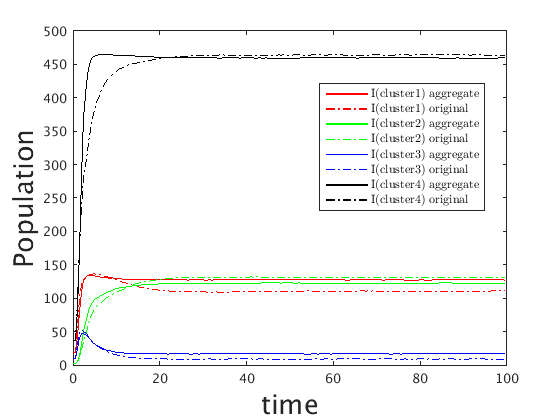}
    \includegraphics[width=0.45\textwidth]{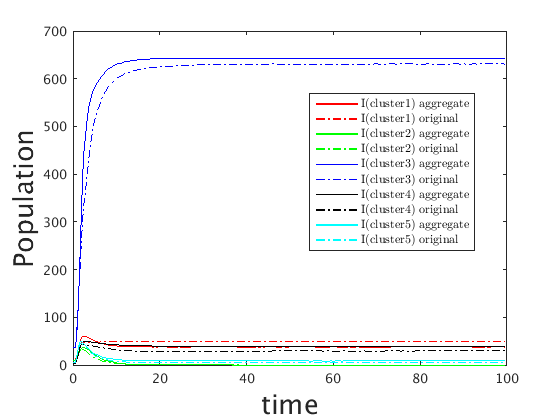}
    \caption{(left) Comparison of the infected populations before and after aggregation for the SIS model, using the $d_L$ distance. (right) Comparison of the infected populations before and after aggregation for the SIS model, using the $d_E$ distance.}
    \label{fig:errprcompare}
\end{figure}

In our experiment, we consider a model with $m=30$, $ \beta_i$ and $r_{ij}$ to be random values between zero and one, $\mu=0.1$. We first report the analysis using the Linear-Noise metric $d_L$. Computing the distance and running the spectral clustering algorithm, according to the previous section, we obtained the spectre of the normalized Laplacian matrix shown in Figure \ref{fig:eigenvaluesSIS} (left), for  the smallest 10 eigenvalues. As can be seen from the figure, the first four eigenvalues are very close to zero, and there is a large gap between the 4th and 5th eigenvalues. Thus, we set the number of clusters to four. Figure~\ref{fig:errprcompare} (left) shows the trajectories of the infected population generated by stochastic simulation before and after aggregation. Table~\ref{tab:siscompare} compares the number of transitions, simulation time of 1000 runs of the SIS model before and after aggregation, as well as the average error ratio of the trajectories in Figure~\ref{fig:errprcompare} (left) after aggregation compared with the counterpart before aggregation. As can be readily seen, our method considerably reduced the simulation cost of the model, at the price of a reasonably small loss of accuracy, the relative error being less than $10\%$ (the average error ratio is computed as the average along the trajectory).
In Figure \ref{fig:errprcompare} (right), instead, we show the result of the aggregation when using the mean-field distance $d_E$. In this case, the spectral clustering identifies 5 clusters, with a comparable overall accuracy with respect to the linear-noise distance as can be seen in Table~\ref{tab:siscompare}.

\begin{table}[t]
\begin{center}
{%
\begin{tabular}{ |c |c |c| c| }
\hline
SIS model & No. of transitions & simulation time (1000 runs) & Avg error ratio  \\
\hline
Before aggregation  & 240 & 8.81 mins & N/A \\
\hline  
After aggregation ($d_L$) & 28 & 59 secs & 8.67\% \\
\hline  
After aggregation ($d_E$) & 34 & 3.67 mins & 10.44\% \\
\hline  
\end{tabular}}
\caption{Size, simulation cost (including the aggregation cost) of the SIS model before and after aggregation,  and error introduced by the aggregation.\label{tab:siscompare}}
\end{center}
\end{table}

\subsection{Public Bike-sharing Model}
\label{sec:BS}
The second example is a spatial PCTMC which models a public bike-sharing system. Bike-sharing systems are becoming more and more important for urban transportation. In such systems, users arrive at a station, pick up a bike, use it for a while, and then return it to another station of their choice. Recently, PCTMCs have been used to model bike-sharing systems \cite{fricker2014incentives,guenther2013journey}. Here, we consider a map which consists of $N$ zones. There is one public bike station in each zone. Each station has several bike slots. The pickup rate of bikes in a station is governed by an exponential distribution. When a user picks up a bike, the available number of bikes in the station will decrease by one whereas the available number of slots in that station will increase by one. The user will choose another zone in the city as their destination. When the user arrives at the destination zone, they will return their bike to the bike station in that zone.  We use a spatial PCTMC containing the following transitions to represent the model:
\begin{small}
\begin{eqnarray*}
Bike(\loc_i) &\rightarrow & Slot(\loc_i) + BikeTo_j(\loc_i) \quad \text{at $\lambda_i \: p^i_j$} \\
BikeTo_j(\loc_i) + Slot(\loc_j) &\rightarrow & Bike(\loc_j) \quad \quad \quad \quad \quad\quad  \: \: \:  \text{at $\#(BikeTo_j(\loc_i)) \: \mu^i_j$} 
\end{eqnarray*}
\end{small}where $Bike(\loc_i)$ and $Slot(\loc_i) $ denote an available bike or slot in the bike station in Zone $i$, respectively; $BikeTo_j(\loc_i)$ denotes a bike in transit from Zone $i$ to Zone $j$. $\lambda_i$ is the pickup rate of bikes in the bike station in Zone $i$, $p^i_j$ is the probability to choose Zone $j$ as the destination of a trip when picking up a bike from Zone $i$. $1/ \mu^i_j$ is the mean trip time from Zone $i$ to Zone $j$.  

\begin{figure}
  \centering
    \includegraphics[width=0.59\textwidth]{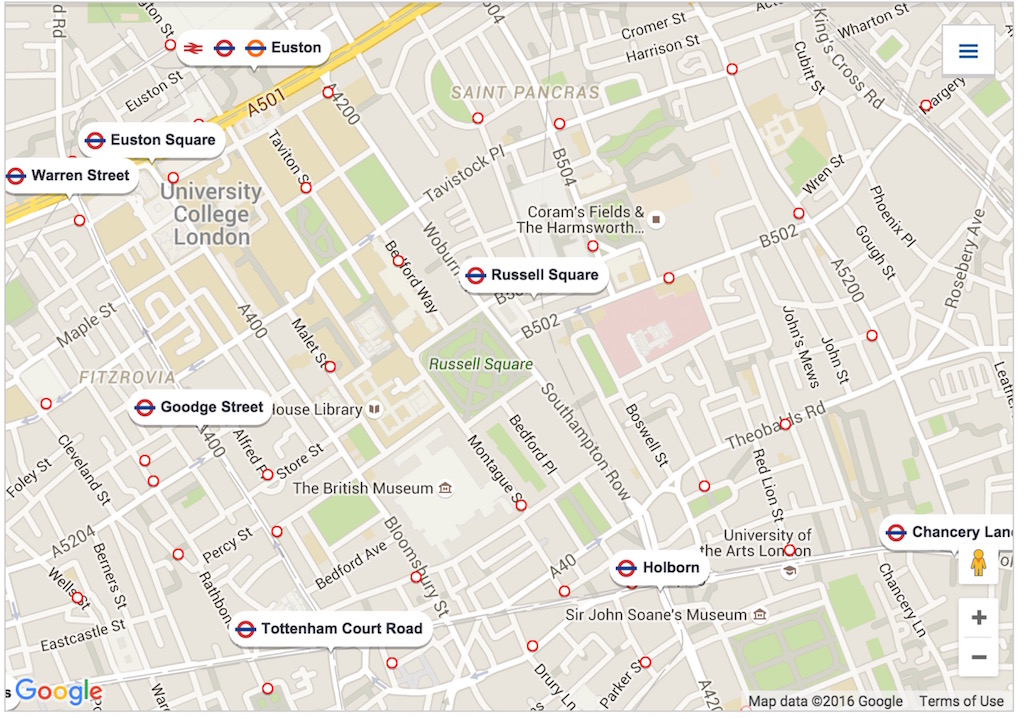}
    \caption{The map of bike-sharing stations near Russell Square in London in which red circles representing bike stations}
    \label{fig:bikemap}
\end{figure}

Specifically, we use the above spatial PCTMC to model 30 bike stations near Russell Square in London which is illustrated in Figure~\ref{fig:bikemap}. All the rates in the model are calculated by journey data which is available online \footnote{\url{https://tfl.gov.uk/info-for/open-data-users/our-feeds?intcmp=3671##on-this-page-4}}.

\begin{figure}[H]
  \centering
    \includegraphics[width=0.45\textwidth]{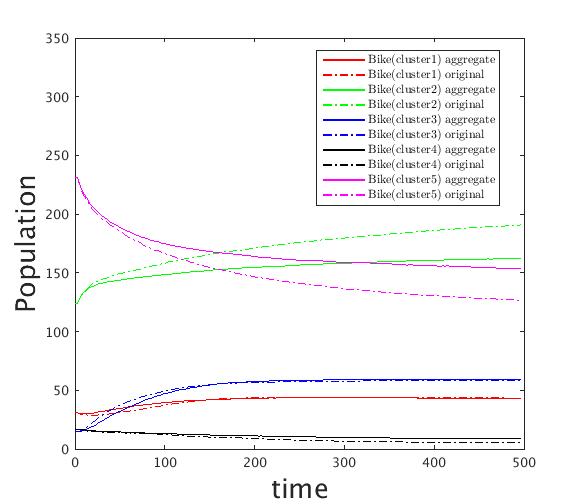}
     \includegraphics[width=0.45\textwidth]{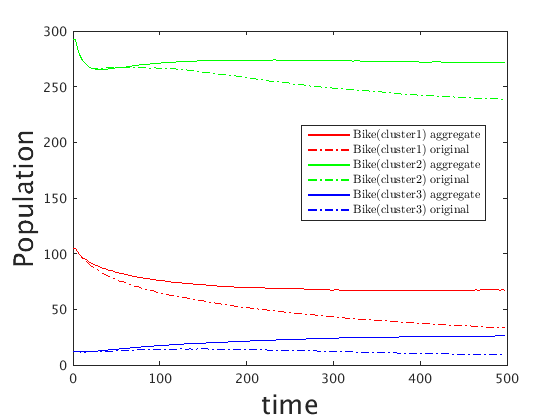}
    \caption{(left) Comparison of the available number of bikes before and after aggregation using the $d_L$ distance. (right) Comparison of the available number of bikes before and after aggregation using the $d_E$ distance.}
    \label{fig:fiveclusters}
\end{figure}

We apply our method with the linear noise distance to cluster the 30 bike stations. The smallest 10 eigenvalues of the normalized Laplacian matrix, computed according to the spectral clustering recipe, are shown in Figure~\ref{fig:eigenvalues} (right). According to the eigengap heuristic, there are 5 well separated clusters. Figure~\ref{fig:fiveclusters} (left) shows the trajectories of the number of available bikes in the 5 clusters generated by stochastic simulation before and after aggregation. Table~\ref{tab:bikecompare} shows the number of transitions, simulation time of 1000 runs of the bike-sharing model before and after aggregation, as well as the average error ratio of the trajectories in Figure~\ref{fig:fiveclusters} (left) after aggregation compared with the counterpart before aggregation. In Figure~\ref{fig:fiveclusters} (right), instead, we compare the trajectories of the original model and the reduced one according to the mean-field metric $d_E$. In this case, we have only three clusters and the accuracy decreases considerably, as can be numerically seen in Table~\ref{tab:bikecompare}.

\begin{table}[t]
\begin{center}
{%
\begin{tabular}{ |c |c |c| c| }
\hline
Bike model & No. of transitions & simulation time (1000 runs) & Avg error ratio  \\
\hline
Before aggregation  & 1800 & 15.6 mins & N/A \\
\hline  
After aggregation ($d_{L}$) & 300 & 2.5 mins & 11.94\% \\
\hline  
After aggregation ($d_{E}$) & 180 & 1.6 mins & 23.06\% \\
\hline  
\end{tabular}}
\caption{Size, simulation cost (including the aggregation cost) of the bike-sharing model before and after aggregation,  and error introduced by the aggregation. \label{tab:bikecompare}}
\end{center}
\end{table}

\begin{figure}[t]
  \centering
    \includegraphics[width=0.45\textwidth]{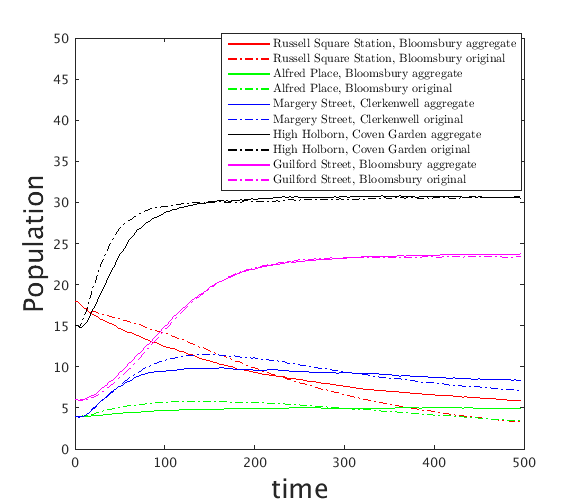}
    \caption{Comparison of the trajectories of the number of available bikes in the five chosen stations from the original and the aggregated model.}
    \label{fig:singlecompare}
\end{figure}

In a bike-sharing scenario, we are often interested in tracking the number of bikes in some specific locations. This can be achieved in our framework by forcing some locations to be a singleton cluster. In order to understand the influence of the aggregation of remaining stations on some isolated ones,  we choose, in each experiment, one station from one of the five clusters, and treat that station as a single cluster. Figure~\ref{fig:singlecompare} shows the trajectories of the number of available bikes of  the five chosen stations in five different experiments,  comparing results of  stochastic simulation of the original and the aggregated model. We can see the population dynamics of available bikes in the five stations still achieve good accuracy ($10.84\%$ average error ratio) even if other bike stations are aggregated. 

\begin{figure}[H]
  \centering
    \includegraphics[width=0.45\textwidth]{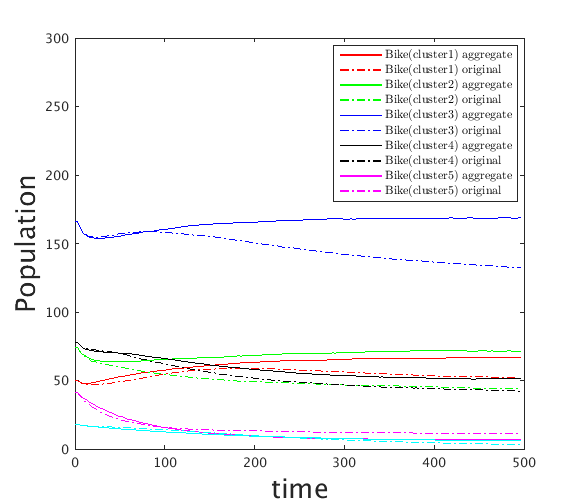}
    \caption{Comparison of the available number of bikes before and after aggregation according to physical position of stations}
    \label{fig:fiveclustersphysical}
\end{figure}

Additionally, we also apply a naive approach in which we use spectral clustering to cluster the 30 bike stations according to their physical positions on the map. Figure~\ref{fig:fiveclustersphysical} shows the trajectories of the number of available bikes generated by stochastic simulation before and after aggregation in this case. The average error ratio for the trajectories is $24.51\%$. Thus,  location clustering according to agents' population distribution outperforms clustering according to physical positions.

\section{Related Work and Conclusion}
\label{sec:conclusion}
Spatiality has been shown to be an important factor for modelling different classes of  complex systems \cite{durrett1994importance,durrett1999stochastic}. As a result, there has been some effort to study the effects of spatial aggregation on stochastic models in order to achieve a balance point between model complexity and accuracy. Most of this work focuses on how to discretise continuous space or on the effect of compartment size to represent the space, see for example \cite{CFeng14,erban2009stochastic,bonachela2012patchiness}. Our work can be thought of as above theses approaches, as our goal is to reduce the number of patches (locations). We show that by clustering locations in a spatial PCTMC according to linear noise or mean field distance, the size of the model can be significantly reduced but still retain reasonable accuracy. The experiment on the bike-sharing model shows that our method outperforms aggregation of locations by their physical distance. Moreover, our recent work shows that the cost of deriving higher moments based on fluid approximation can be significantly reduced using correlation heuristics \cite{cfeng2016tomacs}. This means that computing linear noise distance can also be largely reduced, which makes our method more scalable. In the future, we plan to apply this approach to models in spatial process algebras, such as PALOMA \cite{feng2014paloma}, CARMA \cite{BortolussiNGGHL15} and MELA \cite{mela}.

\section{Acknowledgments}
The authors would like to thank Jane Hillston for her helpful comments on an earlier draft of this work. This work is supported by the EU project QUANTICOL, 600708.

\bibliographystyle{eptcs}
\bibliography{qapl}

\begin{thebibliography}{10}
\providecommand{\bibitemdeclare}[2]{}
\providecommand{\surnamestart}{}
\providecommand{\surnameend}{}
\providecommand{\urlprefix}{Available at }
\providecommand{\url}[1]{\texttt{#1}}
\providecommand{\href}[2]{\texttt{#2}}
\providecommand{\urlalt}[2]{\href{#1}{#2}}
\providecommand{\doi}[1]{doi:\urlalt{http://dx.doi.org/#1}{#1}}
\providecommand{\bibinfo}[2]{#2}

\bibitemdeclare{book}{allen2010introduction}
\bibitem{allen2010introduction}
\bibinfo{author}{Linda~JS \surnamestart Allen\surnameend}
  (\bibinfo{year}{2010}): \emph{\bibinfo{title}{An introduction to stochastic
  processes with applications to biology}}.
\newblock \bibinfo{publisher}{CRC Press}.

\bibitemdeclare{book}{andersson2000}
\bibitem{andersson2000}
\bibinfo{author}{H.~\surnamestart Andersson\surnameend} \&
  \bibinfo{author}{T.~\surnamestart Britton\surnameend} (\bibinfo{year}{2000}):
  \emph{\bibinfo{title}{Stochastic {Epidemic} {Models} and {Their}
  {Statistical} {Analysis}}}.
\newblock \bibinfo{publisher}{Springer-Verlag},
  \doi{10.1007/978-1-4612-1158-7}.

\bibitemdeclare{article}{andreychenko2015}
\bibitem{andreychenko2015}
\bibinfo{author}{A.~\surnamestart Andreychenko\surnameend},
  \bibinfo{author}{L.~\surnamestart Mikeev\surnameend} \&
  \bibinfo{author}{C.~\surnamestart Wolf\surnameend} (\bibinfo{year}{2015}):
  \emph{\bibinfo{title}{Model Reconstruction for Moment-Based Stochastic
  Chemical Kinetics}}.
\newblock {\sl \bibinfo{journal}{ACM Transactions on Modeling and Computer
  Simulation (TOMACS)}} \bibinfo{volume}{25}(\bibinfo{number}{2}),
  p.~\bibinfo{pages}{12}, \doi{10.1145/2699712}.

\bibitemdeclare{article}{bhattachayya1943measure}
\bibitem{bhattachayya1943measure}
\bibinfo{author}{A~\surnamestart Bhattacharyya\surnameend}
  (\bibinfo{year}{1943}): \emph{\bibinfo{title}{On a measure of divergence
  between two statistical population defined by their population
  distributions}}.
\newblock {\sl \bibinfo{journal}{Bulletin Calcutta Mathematical Society}}
  \bibinfo{volume}{35}, pp. \bibinfo{pages}{99--109}.

\bibitemdeclare{article}{bonachela2012patchiness}
\bibitem{bonachela2012patchiness}
\bibinfo{author}{Juan~A \surnamestart Bonachela\surnameend},
  \bibinfo{author}{Miguel~A \surnamestart Mu{\~n}oz\surnameend} \&
  \bibinfo{author}{Simon~A \surnamestart Levin\surnameend}
  (\bibinfo{year}{2012}): \emph{\bibinfo{title}{Patchiness and demographic
  noise in three ecological examples}}.
\newblock {\sl \bibinfo{journal}{Journal of Statistical Physics}}
  \bibinfo{volume}{148}(\bibinfo{number}{4}), pp. \bibinfo{pages}{724--740},
  \doi{10.1007/s10955-012-0506-x}.

\bibitemdeclare{article}{tutorial}
\bibitem{tutorial}
\bibinfo{author}{L.~\surnamestart Bortolussi\surnameend},
  \bibinfo{author}{J.~\surnamestart Hillston\surnameend},
  \bibinfo{author}{D.~\surnamestart Latella\surnameend} \&
  \bibinfo{author}{M.~\surnamestart Massink\surnameend} (\bibinfo{year}{2013}):
  \emph{\bibinfo{title}{Continuous approximation of collective system
  behaviour: A tutorial}}.
\newblock {\sl \bibinfo{journal}{Performance Evaluation}}
  \bibinfo{volume}{70}(\bibinfo{number}{5}), pp. \bibinfo{pages}{317--349},
  \doi{10.1016/j.peva.2013.01.001}.

\bibitemdeclare{inproceedings}{valuetools2014}
\bibitem{valuetools2014}
\bibinfo{author}{L.~\surnamestart Bortolussi\surnameend} \&
  \bibinfo{author}{L.~\surnamestart Nenzi\surnameend} (\bibinfo{year}{2014}):
  \emph{\bibinfo{title}{Specifying and monitoring properties of stochastic
  spatio-temporal systems in signal temporal logic}}.
\newblock In: {\sl \bibinfo{booktitle}{Proceedings of the 8th International
  Conference on Performance Evaluation Methodologies and Tools}}, pp.
  \bibinfo{pages}{66--73}, \doi{10.4108/icst.Valuetools.2014.258183}.

\bibitemdeclare{inproceedings}{BortolussiNGGHL15}
\bibitem{BortolussiNGGHL15}
\bibinfo{author}{Luca \surnamestart Bortolussi\surnameend},
  \bibinfo{author}{Rocco \surnamestart {De Nicola}\surnameend},
  \bibinfo{author}{Vashti \surnamestart Galpin\surnameend},
  \bibinfo{author}{Stephen \surnamestart Gilmore\surnameend},
  \bibinfo{author}{Jane \surnamestart Hillston\surnameend},
  \bibinfo{author}{Diego \surnamestart Latella\surnameend},
  \bibinfo{author}{Michele \surnamestart Loreti\surnameend} \&
  \bibinfo{author}{Mieke \surnamestart Massink\surnameend}
  (\bibinfo{year}{2015}): \emph{\bibinfo{title}{{CARMA:} Collective Adaptive
  Resource-sharing Markovian Agents}}.
\newblock In: {\sl \bibinfo{booktitle}{Proceedings Thirteenth Workshop on
  Quantitative Aspects of Programming Languages and Systems, {QAPL} 2015,
  London, UK.}}, pp. \bibinfo{pages}{16--31}, \doi{10.4204/EPTCS.194.2}.

\bibitemdeclare{incollection}{lanciani2014}
\bibitem{lanciani2014}
\bibinfo{author}{Luca \surnamestart Bortolussi\surnameend} \&
  \bibinfo{author}{Roberta \surnamestart Lanciani\surnameend}
  (\bibinfo{year}{2014}): \emph{\bibinfo{title}{Stochastic approximation of
  global reachability probabilities of Markov population models}}.
\newblock In: {\sl \bibinfo{booktitle}{Computer Performance Engineering}},
  \bibinfo{publisher}{Springer}, pp. \bibinfo{pages}{224--239},
  \doi{10.1007/978-3-319-10885-8\_16}.

\bibitemdeclare{article}{durrett1994importance}
\bibitem{durrett1994importance}
\bibinfo{author}{Richard \surnamestart Durrett\surnameend} \&
  \bibinfo{author}{Simon \surnamestart Levin\surnameend}
  (\bibinfo{year}{1994}): \emph{\bibinfo{title}{The importance of being
  discrete (and spatial)}}.
\newblock {\sl \bibinfo{journal}{Theoretical population biology}}
  \bibinfo{volume}{46}(\bibinfo{number}{3}), pp. \bibinfo{pages}{363--394},
  \doi{10.1006/tpbi.1994.1032}.

\bibitemdeclare{article}{durrett1999stochastic}
\bibitem{durrett1999stochastic}
\bibinfo{author}{Rick \surnamestart Durrett\surnameend} (\bibinfo{year}{1999}):
  \emph{\bibinfo{title}{Stochastic spatial models}}.
\newblock {\sl \bibinfo{journal}{SIAM review}}
  \bibinfo{volume}{41}(\bibinfo{number}{4}), pp. \bibinfo{pages}{677--718},
  \doi{10.1137/S0036144599354707}.

\bibitemdeclare{article}{erban2009stochastic}
\bibitem{erban2009stochastic}
\bibinfo{author}{Radek \surnamestart Erban\surnameend} \&
  \bibinfo{author}{S~Jonathan \surnamestart Chapman\surnameend}
  (\bibinfo{year}{2009}): \emph{\bibinfo{title}{Stochastic modelling of
  reaction--diffusion processes: algorithms for bimolecular reactions}}.
\newblock {\sl \bibinfo{journal}{Physical biology}}
  \bibinfo{volume}{6}(\bibinfo{number}{4}), p. \bibinfo{pages}{046001},
  \doi{10.1088/1478-3975/6/4/046001}.

\bibitemdeclare{inproceedings}{CFeng14}
\bibitem{CFeng14}
\bibinfo{author}{Cheng \surnamestart Feng\surnameend} (\bibinfo{year}{2014}):
  \emph{\bibinfo{title}{Patch-based Hybrid Modelling of Spatially Distributed
  Systems by Using Stochastic {HYPE} - ZebraNet as an Example}}.
\newblock In: {\sl \bibinfo{booktitle}{Proceedings Twelfth International
  Workshop on Quantitative Aspects of Programming Languages and Systems, {QAPL}
  2014, Grenoble, France.}}, pp. \bibinfo{pages}{64--77},
  \doi{10.4204/EPTCS.154.5}.

\bibitemdeclare{incollection}{feng2014paloma}
\bibitem{feng2014paloma}
\bibinfo{author}{Cheng \surnamestart Feng\surnameend} \& \bibinfo{author}{Jane
  \surnamestart Hillston\surnameend} (\bibinfo{year}{2014}):
  \emph{\bibinfo{title}{PALOMA: A process algebra for located markovian
  agents}}.
\newblock In: {\sl \bibinfo{booktitle}{Quantitative Evaluation of Systems}},
  \bibinfo{publisher}{Springer}, pp. \bibinfo{pages}{265--280},
  \doi{10.1007/978-3-319-10696-0\_22}.

\bibitemdeclare{article}{cfeng2016tomacs}
\bibitem{cfeng2016tomacs}
\bibinfo{author}{Cheng \surnamestart Feng\surnameend}, \bibinfo{author}{Jane
  \surnamestart Hillston\surnameend} \& \bibinfo{author}{Vashti \surnamestart
  Galpin\surnameend} (\bibinfo{year}{2016}): \emph{\bibinfo{title}{Automatic
  Moment-Closure Approximation of Spatially Distributed Collective Adaptive
  Systems}}.
\newblock {\sl \bibinfo{journal}{ACM Transactions on Modeling and Computer
  Simulation (TOMACS)}} \bibinfo{volume}{26}(\bibinfo{number}{4}),
  \doi{10.1145/2883608}.

\bibitemdeclare{article}{fricker2014incentives}
\bibitem{fricker2014incentives}
\bibinfo{author}{Christine \surnamestart Fricker\surnameend} \&
  \bibinfo{author}{Nicolas \surnamestart Gast\surnameend}
  (\bibinfo{year}{2014}): \emph{\bibinfo{title}{Incentives and redistribution
  in homogeneous bike-sharing systems with stations of finite capacity}}.
\newblock {\sl \bibinfo{journal}{EURO Journal on Transportation and
  Logistics}}, pp. \bibinfo{pages}{1--31}, \doi{10.1007/s13676-014-0053-5}.

\bibitemdeclare{article}{gillespie1977exact}
\bibitem{gillespie1977exact}
\bibinfo{author}{Daniel~T \surnamestart Gillespie\surnameend}
  (\bibinfo{year}{1977}): \emph{\bibinfo{title}{Exact stochastic simulation of
  coupled chemical reactions}}.
\newblock {\sl \bibinfo{journal}{The Journal of Physical Chemistry}}
  \bibinfo{volume}{81}(\bibinfo{number}{25}), pp. \bibinfo{pages}{2340--2361},
  \doi{10.1021/j100540a008}.

\bibitemdeclare{incollection}{guenther2013journey}
\bibitem{guenther2013journey}
\bibinfo{author}{Marcel~C \surnamestart Guenther\surnameend} \&
  \bibinfo{author}{Jeremy~T \surnamestart Bradley\surnameend}
  (\bibinfo{year}{2013}): \emph{\bibinfo{title}{Journey data based arrival
  forecasting for bicycle hire schemes}}.
\newblock In: {\sl \bibinfo{booktitle}{Analytical and Stochastic Modeling
  Techniques and Applications}}, \bibinfo{publisher}{Springer}, pp.
  \bibinfo{pages}{214--231}, \doi{10.1007/978-3-642-39408-9\_16}.

\bibitemdeclare{incollection}{guenther2013moment}
\bibitem{guenther2013moment}
\bibinfo{author}{Marcel~C \surnamestart Guenther\surnameend},
  \bibinfo{author}{Anton \surnamestart Stefanek\surnameend} \&
  \bibinfo{author}{Jeremy~T \surnamestart Bradley\surnameend}
  (\bibinfo{year}{2013}): \emph{\bibinfo{title}{Moment closures for performance
  models with highly non-linear rates}}.
\newblock In: {\sl \bibinfo{booktitle}{Computer Performance Engineering}},
  \bibinfo{publisher}{Springer}, pp. \bibinfo{pages}{32--47},
  \doi{10.1007/978-3-642-36781-6\_3}.

\bibitemdeclare{article}{hasenauer2013}
\bibitem{hasenauer2013}
\bibinfo{author}{J.~\surnamestart Hasenauer\surnameend},
  \bibinfo{author}{V.~\surnamestart Wolf\surnameend},
  \bibinfo{author}{A.~\surnamestart Kazeroonian\surnameend} \&
  \bibinfo{author}{F.~J. \surnamestart Theis\surnameend}
  (\bibinfo{year}{2013}): \emph{\bibinfo{title}{Method of conditional moments
  ({MCM}) for the {Chemical} {Master} {Equation}: {A} unified framework for the
  method of moments and hybrid stochastic-deterministic models}}.
\newblock {\sl \bibinfo{journal}{Journal of Mathematical Biology}},
  \doi{10.1007/s00285-013-0711-5}.

\bibitemdeclare{book}{vankampen}
\bibitem{vankampen}
\bibinfo{author}{N.~G. \surnamestart van Kampen\surnameend}
  (\bibinfo{year}{2007}): \emph{\bibinfo{title}{Stochastic processes in physics
  and chemistry}}.
\newblock \bibinfo{publisher}{Elsevier}.

\bibitemdeclare{book}{kurtz1981}
\bibitem{kurtz1981}
\bibinfo{author}{T.~G. \surnamestart Kurtz\surnameend} (\bibinfo{year}{1981}):
  \emph{\bibinfo{title}{Approximation of population processes}}.
\newblock \bibinfo{publisher}{SIAM}, \doi{10.1137/1.9781611970333}.

\bibitemdeclare{incollection}{moharbojan}
\bibitem{moharbojan}
\bibinfo{author}{Bojan \surnamestart Mohar\surnameend} (\bibinfo{year}{1997}):
  \emph{\bibinfo{title}{Some applications of Laplace eigenvalues of graphs}}.
\newblock In: {\sl \bibinfo{booktitle}{Graph Symmetry}}, {\sl
  \bibinfo{series}{NATO ASI Series}} \bibinfo{volume}{497},
  \bibinfo{publisher}{Springer Netherlands}, pp. \bibinfo{pages}{225--275},
  \doi{10.1007/978-94-015-8937-6\_6}.

\bibitemdeclare{article}{ng2002spectral}
\bibitem{ng2002spectral}
\bibinfo{author}{Andrew~Y \surnamestart Ng\surnameend},
  \bibinfo{author}{Michael~I \surnamestart Jordan\surnameend},
  \bibinfo{author}{Yair \surnamestart Weiss\surnameend} et~al.
  (\bibinfo{year}{2002}): \emph{\bibinfo{title}{On spectral clustering:
  Analysis and an algorithm}}.
\newblock {\sl \bibinfo{journal}{Advances in neural information processing
  systems}} \bibinfo{volume}{2}, pp. \bibinfo{pages}{849--856},
  \doi{10.1.1.19.8100}.

\bibitemdeclare{article}{shi2000normalized}
\bibitem{shi2000normalized}
\bibinfo{author}{Jianbo \surnamestart Shi\surnameend} \&
  \bibinfo{author}{Jitendra \surnamestart Malik\surnameend}
  (\bibinfo{year}{2000}): \emph{\bibinfo{title}{Normalized cuts and image
  segmentation}}.
\newblock {\sl \bibinfo{journal}{IEEE Transactions on Pattern Analysis and
  Machine Intelligence}} \bibinfo{volume}{22}(\bibinfo{number}{8}), pp.
  \bibinfo{pages}{888--905}, \doi{10.1109/34.868688}.

\bibitemdeclare{article}{stefanek2011fluid}
\bibitem{stefanek2011fluid}
\bibinfo{author}{Anton \surnamestart Stefanek\surnameend},
  \bibinfo{author}{Richard~A \surnamestart Hayden\surnameend} \&
  \bibinfo{author}{Jeremy~T \surnamestart Bradley\surnameend}
  (\bibinfo{year}{2011}): \emph{\bibinfo{title}{Fluid computation of the
  performance: energy tradeoff in large scale {Markov} models}}.
\newblock {\sl \bibinfo{journal}{ACM SIGMETRICS Performance Evaluation Review}}
  \bibinfo{volume}{39}(\bibinfo{number}{3}), pp. \bibinfo{pages}{104--106},
  \doi{10.1145/2160803.2160817}.

\bibitemdeclare{unpublished}{mela}
\bibitem{mela}
\bibinfo{author}{Ludovica~Luisa \surnamestart Vissat\surnameend},
  \bibinfo{author}{Jane \surnamestart Hillston\surnameend},
  \bibinfo{author}{Glenn \surnamestart Marion\surnameend} \&
  \bibinfo{author}{Matthew~J \surnamestart Smith\surnameend} (\bibinfo{year}{to
  appear}): \emph{\bibinfo{title}{{MELA}: Modelling in Ecology with Location
  Attributes}}.
\newblock \bibinfo{note}{Proceedings of the fourth international symposium on
  Modelling and Knowledge Management applications: Systems and Domains,
  {MoKMaSD} 2015, York, UK}.

\bibitemdeclare{article}{von2007tutorial}
\bibitem{von2007tutorial}
\bibinfo{author}{Ulrike \surnamestart Von~Luxburg\surnameend}
  (\bibinfo{year}{2007}): \emph{\bibinfo{title}{A tutorial on spectral
  clustering}}.
\newblock {\sl \bibinfo{journal}{Statistics and computing}}
  \bibinfo{volume}{17}(\bibinfo{number}{4}), pp. \bibinfo{pages}{395--416},
  \doi{10.1007/s11222-007-9033-z}.

\end{thebibliography}
\end{document}